\documentclass{article}
\pdfoutput=1
\usepackage[a4paper]{geometry}
\usepackage[T1]{fontenc}
\usepackage[utf8]{inputenc}
\usepackage{lmodern}

\usepackage{graphicx}

\usepackage{amsmath,amssymb}
\usepackage{amsthm}
\usepackage{subcaption}
\usepackage{algorithm}
\usepackage{algorithmicx,algpseudocode}
\usepackage{adjustbox}
\usepackage{siunitx}

\usepackage{xcolor}

\definecolor{col1}{RGB}{100,143,255}
\definecolor{col2}{RGB}{120, 94, 240}
\definecolor{col3}{RGB}{254,97,0}
\definecolor{col4}{RGB}{220, 38, 127}
\definecolor{col5}{RGB}{255, 176, 0}

\usepackage{hyperref}

\newtheorem{theorem}{Theorem}%

\newtheorem{conjecture}[theorem]{Conjecture}

\theoremstyle{remark}

\newcommand{\dd}{\text{d}}

\newcommand{\bigO}{\mathcal{O}}

\DeclareMathOperator{\Aut}{Aut}

\hypersetup{
  pdfauthor={Michael Borinsky, Andrea Favorito},
  pdftitle={Feynman integrals at large loop order and the log-Gamma distribution}
  linkcolor  = col2,
  citecolor  = col4,
  urlcolor   = col1,
  colorlinks = true,
}

\definecolor{red}{rgb}{1,0,0}

 \definecolor{darkgreen}{rgb}{0, .7, 0}

 \definecolor{purple}{rgb}{.7, 0, 1}

\usepackage[normalem]{ulem}

\begin{document}

\title{Feynman integrals at large loop order 
\\
and the $\log$-$\Gamma$ distribution
}
\author{Michael Borinsky\thanks{
Perimeter Institute, Waterloo, ON N2L 2Y5, Canada
} 
$^,$\thanks{
Institute for Theoretical Studies, ETH Zürich,
8092 Zürich, Switzerland
}
 \and Andrea Favorito\thanks{
Institute for Theoretical Physics, ETH Zürich, 8093 Zürich, Switzerland
}
}
\date{}
\maketitle

\begin{abstract}
We find empirically that the value of Feynman integrals follows a $\log$-$\Gamma$ distribution at large loop order.
This result opens up a new avenue towards the large-order behavior in perturbative quantum field theory.
 Our study of the primitive contribution to the scalar $\phi^4$ beta function in four dimensions up to 17 loops provides accompanying evidence. 
Guided by instanton considerations, we discuss the extrapolation of this contribution to all loop orders.
\end{abstract}
\section{Introduction}

\emph{Feynman integrals} are the building blocks of perturbative quantum, statistical, and classical field theory expansions. Each Feynman integral corresponds to a Feynman graph and contributes to a specific perturbative order. In most cases, this order equals the graph's loop number. A perturbative expansion is a formal power series $ \mathcal A(\hbar) = \sum_{L \geq 0} A_L \hbar^L$,
with the $L$-th coefficient given by a sum of Feynman graphs:
\begin{align} \label{eq:AL} A_L = \sum_{L(G) = L} \frac{I_G}{|\Aut(G)|}, \end{align}
where $\hbar$ is the perturbative expansion parameter, we sum over all Feynman graphs $G$ of specific shape and loop order $L(G)$, $I_G$ is the Feynman integral corresponding to the graph,
and $|\Aut(G)|$ denotes the graph's \emph{symmetry factor} 
(i.e.~the order of its \emph{automorphism group}).
The precise shape of the graphs and the associated integral depend  on the specific underlying theory.

The type of perturbative expansion~\eqref{eq:AL} allows the prediction of a large variety of physical phenomena. For instance, the \emph{Feynman amplitude} in various quantum field theories is typically expanded via \eqref{eq:AL} (see, e.g., \cite{Travaglini:2022uwo}). The \emph{critical exponents} of various interesting universality classes can be computed from similar expressions (e.g., for the $3$-dimensional Ising model and $3$-dimensional percolation theory \cite{Wilson:1971dc}). Even general relativity corrections to the Newton potential can be computed using Feynman integral sums as the one above (see, e.g., \cite{Kosower:2018adc}).

Much progress has been made in computing the coefficients $A_L$ in recent decades (e.g., \cite{Chetyrkin:1981qh,Laporta:2000dsw,Henn:2013pwa,Panzer:2014caa}). Improvements in understanding the underlying mathematical structures of amplitudes, Feynman integrals, and their singularities enabled these leaps \cite{Britto:2005fq,Bloch:2005bh,Bern:2007dw,Brown:2008um}.

Regarding the  ubiquitous nature of Feynman integrals and sums, we ask:
What is the distribution of the values contributing to~\eqref{eq:AL}?
Are all $I_G$ of the same magnitude, or do particular graphs contribute more? 
We thus study the \emph{distribution of the value of Feynman integrals}.

Ultimately, this question is motivated by the \emph{perturbation theory at large order} program \cite{Bender:1969si,le2012large} and its recent incarnation, the \emph{resurgence program} (see, e.g., \cite{Dorigoni:2014hea}). The progress from both these programs suggests that the large-$L$ behavior of the coefficients $A_L$ encodes much (and perhaps all) \emph{nonperturbative} information of the function $\mathcal A(\hbar)$. %
Further, an ongoing research program aims to replace sums as~\eqref{eq:AL} with integrals over a single, continuous object \cite{Arkani-Hamed:2010zjl,Cachazo:2013hca,Arkani-Hamed:2017tmz,Arkani-Hamed:2023lbd}.
Our results in Sec.~\ref{sec:histos} provide concrete information on such an object's expected \emph{asymptotic} $L\rightarrow \infty$ shape.

This article will focus on a specific model: $\phi^4$ quantum field theory in $D = 4 - 2 \varepsilon$ dimensions. %
Within this model, we focus on a particular subset of terms contributing to the beta function: the \emph{primitive} contribution.
The $\phi^4$ beta function is known exactly up to loop order $7$  in the minimal subtraction (MS) scheme \cite{Schnetz:2022nsc,Batkovich:2016jus,Kleinert:1991rg,Chetyrkin:1981qh}. The primitive contribution to this beta function is obtained by summing over all \emph{period Feynman integrals}, given by the $1/\varepsilon$ residues of specific Feynman integrals that are free of subdivergences (see \S\ref{sec:method} for a precise definition of period Feynman integrals). It is conjectured  that asymptotically, at large loop order,  the primitive contribution gives the full beta function of $\phi^4$ theory in the MS scheme \cite{McKane:1984eq} and other contributions are subdominant in that limit (see also \cite{Kompaniets_2017,McKane:2018ocs,Dunne:2021lie}). 

In \S \ref{sec:method}, we use an \emph{empirical}, \emph{numerical} approach to study the terms in the sum \eqref{eq:AL} when $L$ is large.  At sufficiently large loop order, exact computation methods for general Feynman integrals will inevitably fail. %
Here, we use the \emph{tropical sampling approach} introduced by the first author in \cite{Borinsky:2020rqs} to evaluate many Feynman integrals with up to 17 loops (see also \cite{Borinsky:2023jdv}). The tropical sampling method draws from previous ideas of \emph{sector decomposition} \cite{Binoth:2003ak,Kaneko:2009qx}, the \emph{Hepp bound}~\cite{Panzer:2019yxl}, and the algebraic geometry of Feynman integrals \cite{Bloch:2005bh,Brown:2015fyf}, which it leverages with properties of generalized permutahedra~\cite{aguiar2023hopf}.
Balduf and Balduf--Shaban recently performed similar large-scale computations of Feynman integrals using the tropical approach \cite{Balduf:2023ilc,Balduf:2024gvv}. 
Balduf estimated the 
 primitive contribution to the $\phi^4$-theory beta function up to 18 loops~\cite{Balduf:2023ilc}.  We confirm these computations up to~17~loops. 

The focal point of this article is the \emph{distribution} of the values in the sum~\eqref{eq:AL}. Our main result is  such a (conjectured) limiting distribution of period Feynman integrals in $\phi^4$ theory for large $L$. %
The simplicity of our result suggests similar structures within other observables in more elaborate quantum field theories.
The following histogram illustrates this limiting distribution:
\begin{figure}[h!]
  \centering
  \includegraphics[]{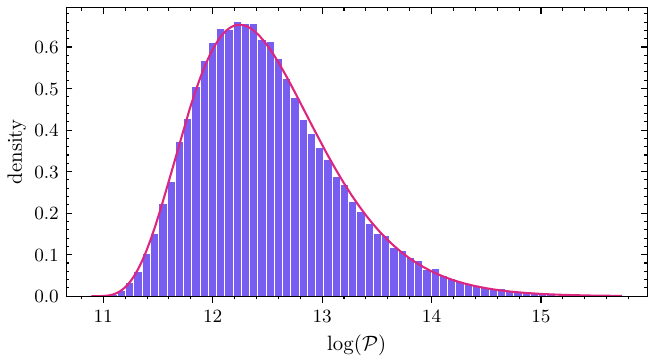}
  \caption{Distribution of $\phi^4$ period Feynman integrals at 17 loops.} 
  \label{fig:hist1}
\end{figure}

The histogram shows our measured distribution of $\phi^4$ period Feynman integrals at 17 loops. 
We obtained this distribution by evaluating $44027$ Feynman integrals up to $10^{-3}$ relative accuracy.  Histograms similar to Figure~\ref{fig:hist1} already appeared in \cite{BENDER1978250} (see also \cite{Bender:1976ni}), which studied the distribution of Feynman integrals with a related underlying  motivation. Here, thanks to more advanced tools, we have the advantage of being able to probe a much higher order in perturbation theory. %
Below, Figure~\ref{fig:hist2} shows similar histograms for lower loop orders and illustrates the rapid convergence to the limiting distribution.
The red curve depicts the density
\begin{align} \label{eq:loggamma} \mu(\mathcal P)= \frac{\lambda^\alpha}{\Gamma(\alpha)} \left(\log \frac{\mathcal P}{\mathcal P_0}\right)^{\alpha-1} \left(\frac{\mathcal P}{\mathcal P_0}\right)^{-\lambda} \dd( \log \mathcal P), \end{align}
where $\mu(\mathcal P)$ is the density of period Feynman integrals with value $\mathcal P$, and the parameters 
$\alpha=6.41(5)$, $\lambda=3.87(3)$, and ${{\mathcal P_0} = 5.14(7) \cdot 10^4}$ 
are fitted and specific to $L=17$ (see \S \ref{sec:histos} for details on the fitting method). The density~\eqref{eq:loggamma} is called a $\log$-$\Gamma$ distribution (see, e.g., \cite[Example~18.4]{klugman2012loss}) because, as a function of $\log \mathcal P$, it is the shifted integrand of Euler's representation for the $\Gamma$ function,
$ \Gamma(z) = \int_0^\infty t^{z-1} e^{-t} \dd t. $
This motivates
\begin{conjecture}
\label{conj}
The distribution of period Feynman integrals in $\phi^4$ theory at $L$ loops sampled with a probability inversely proportional to the symmetry factor of the associated Feynman graph weakly converges to a $\log$-$\Gamma$ distribution when $L\rightarrow \infty$.
\end{conjecture}

We expect this conjecture to hold for broader classes of Feynman integrals than just period Feynman integrals in $\phi^4$ theory.  For instance, the extrapolation of the findings of \cite{BENDER1978250} by analogy to our results suggests that similar limiting distributions can be observed in QED.

In general, the coefficients $A_L$ are expected to grow factorially \cite{Dyson:1952tj} and, hence, 
the perturbative series, such as 
$ \mathcal A(\hbar) = \sum_{L \geq 0} A_L \hbar^L$,
to be divergent if $\hbar \neq 0$. 
To still utilize the information of the coefficients $A_L$ and 
to rigorously compute a quantity such as $\mathcal A(\hbar)$ for nonzero $\hbar$,
\emph{resummation techniques} are needed. 
As input data, these techniques need information about the $L\rightarrow \infty$ asymptotic behavior of $A_L$. For this reason, this asymptotic behavior is of high conceptual and practical value for the perturbative quantum field theory framework \cite{Parisi:1976zh,brezin1978critical,LeGuillou:1979ixc,Costin:2020pcj,Giorgini:2024aer}. 
One concrete application is the prediction of critical exponents of various interesting universality classes via the Wilson--Fisher approach \cite{Wilson:1971dc} (see also \cite{Brezin:2023kyv} for a recent discussion of the relevance of large-order contributions in this domain).
 In \S \ref{sec:res}, we tabulate our 17 loop order results 
for the value of the primitive contribution to the $\phi^4$ beta function.  We then extrapolate these values to the limit $L\rightarrow \infty$ and discuss the relation to classic conjectures on the asymptotic growth rate of the $\phi^4$  beta function in the MS scheme~(e.g.~to~\cite{Lipatov:1976ny,McKane:1984eq,McKane:2018ocs,Dunne:2021lie}).
We conclude in \S \ref{sec:conc}.

\section{Methodology}
\label{sec:method}
\subsection{The primitive contribution to the \texorpdfstring{$\phi^4$}{phi4} beta function}
In four dimensions, the superficial degree of divergence of a scalar Feynman graph is given by 
$\omega(G) = |E_G| - 2 L(G)$, where $|E_G|$ is the number of edges of the graph and $L(G)$ its loop number~\cite{Weinberg:1959nj}.
A graph $G$ is \emph{primitive divergent} in $\phi^4$ theory if it is 1PI, and $\omega(\gamma) > 0$ for each proper subgraph $\gamma \subsetneq G$, while $\omega(G) =0$. The last conditions ensure the graph has an overall logarithmic divergence and no subdivergences in four-dimensional spacetime. As usual, we will consider the legs of the graphs \emph{fixed} or equivalently \emph{distinguishable}. The condition $\omega(G) =0$ in $\phi^4$ theory in four dimensions implies that the graph has precisely four external legs.

The momentum representation of an $L$-loop Feynman integral in $D$ dimensions reads
\begin{align} I_G = \frac{1}{\pi^{LD/2}} \int \frac{\dd^D k_1 \cdots \dd^D k_L}{\prod_{e\in E_G} Q_e}, \end{align}
where $Q_e = q_e^2 -m_e^2+i0$ is the Feynman propagator associated with an edge $e$, and we integrate over $L$ copies of Minkowski space.
We will assume that the external kinematics are sufficiently generic, so there are no IR divergences.

If $G$ has $L$ loops, $G$ is primitive divergent in a QFT that is renormalizable in four dimensions and $D=4-2\varepsilon$,
then $I_G$, as a function of $\varepsilon$, has a simple pole at $\varepsilon=0$. The value of the associated residue is independent of the external kinematics associated to the graph, i.e.,
\begin{align} I_G = \frac{ \mathcal P(G) }{\varepsilon L } + \bigO(\varepsilon^0) \text{ as } \varepsilon \rightarrow 0, \end{align}
and $\mathcal P(G)$ is a number independent of the external kinematics. We call the number $\mathcal P(G)$ the \emph{period Feynman integral} associated with graph $G$ (see, e.g., \cite{Borinsky:2022lds} for more details on periods).

In this article, we focus on the primitive contribution to the $\phi^4$ beta function (see \cite[Appendix B]{Kompaniets_2017}, whose notation we follow, for details).
This contribution is renormalization scheme independent.
We may express this contribution in terms of period Feynman integrals,\footnote{
The convential shift by one in $\beta^{\mathrm{prim}}_{L+1}$
is due to the fact that $L$-loop vertex diagrams contribute to the $g^{L+1}$ coefficient 
of the beta function in the coupling $g$.
}
\begin{align} \label{eq:betaprim} \beta^{\mathrm{prim}}_{L+1} = 2 \sum_{ \substack{G\\
\phi^4 \text{ primitive}\\
L(G) = L }} \frac{\mathcal P(G)}{|\Aut G|}, \end{align}
where we sum over all primitive divergent $\phi^4$ graphs at $L$ loops. 

We aim to get a clearer picture of the behavior of this sum and its terms when $L$ is large. 
For $L=17$, the sum \eqref{eq:betaprim} has $\approx 7 \cdot 10^{12}$ terms, making it impractical to evaluate all of them individually. For this reason, we use a \emph{sampling} approach to study \eqref{eq:betaprim} and its~terms.

\subsection{A probabilistic approach to the sum over graphs}

Instead of summing over the potentially large number of Feynman graphs at fixed loop order $L$, we will sample such graphs $G$ with probability 
\begin{align} \label{eq:pgraphs} p(G) = \frac{1}{Z_L} \frac{1}{|\Aut G|}, \end{align}
where the normalization factor $Z_L$ is given by 
$$
Z_L = 
\sum_{
\substack{G\\
\phi^4 \text{ primitive}\\
L(G) = L
}} \frac{1}{|\Aut G|}.
$$
For our purposes, it is convenient to sample graphs using this non-uniform distribution.  One reason is that it fits naturally with the perturbative expansion in Eq.~\eqref{eq:AL}, where each graph is weighted by $1/|\Aut(G)|$. Further, it is relatively straightforward to implement a sampling algorithm for this distribution, as we will show below. Moreover, we avoid the cumbersome explicit computation of the cardinality of the automorphism group $\Aut(G)$ (see, e.g., \cite{mckay2008graph}) entirely.

The following algorithm generates samples of primitive divergent graphs with probability \eqref{eq:pgraphs}:
\begin{algorithm}[H]
    \caption{Generate random primitive $L$-loop $\phi^4$ graph}
    \label{algo}
    \begin{algorithmic}[1] %
\State
Start with $L+1$ isolated vertices that each have four distinguishable legs.
\State
Randomly select two of the legs and connect them, replacing two legs with one new edge.
\State
Repeat the last step until only four legs are left.
\State
If the resulting graph is primitive divergent, then return the graph.  If not, go back to step 1.
    \end{algorithmic}
\end{algorithm}

As the legs are distinguishable (which can be realized on the computer by numbering them, for example), the graph returned by the algorithm will have four distinguishable legs.
It follows from a simple combinatorial argument and the \emph{orbit stabilizer theorem} (see, e.g., Lemma 1 of \cite[\S 5.1]{yeats2017combinatorial})
that the algorithm above randomly produces graphs $G$ with probability $p(G)$ from~\eqref{eq:pgraphs}.
Combining \eqref{eq:betaprim} and \eqref{eq:pgraphs}, gives
\begin{equation}
\label{eq:beta_L}
  \beta^{\mathrm{prim}}_{L+1} = 
2 \cdot Z_L \cdot
\sum_{
\substack{G\\
\phi^4 \text{ primitive}\\
L(G) = L
}}
p(G) \cdot \mathcal P(G)
=
2 \cdot Z_L \cdot 
  \langle \mathcal P(G) \rangle_L,
\end{equation}
where we recover the expectation value 
  $\langle \mathcal P(G) \rangle_L$
of the random variable $\mathcal P(G)$ under the 
probabilistic process over  the set of $L$-loop primitive divergent graphs described by Algorithm~\ref{algo}.

We, therefore, can study the sum \eqref{eq:betaprim} by sampling graphs using this algorithm. We do so as follows:
We first generate a sample of a primitive divergent graph using Algorithm~\ref{algo}. Then, we evaluate the associated period Feynman integral using the tropical sampling algorithm from \cite{Borinsky:2020rqs} and the implementation \cite{githubtropsampling} (see the second author's Master's thesis \cite{andrea} for a gentle introduction to the tropical integration method). We configured this tropical sampling algorithm to compute the value of each sampled Feynman graph to about $10^{-3}$ relative accuracy. The resulting number provides one data point for an evaluated period Feynman integral. 
We repeat these steps a large number of times. 
For example, Figure~\ref{fig:hist1} summarizes all 44027 data points we obtained at 17 loops by
running Algorithm~\ref{algo} and the tropical sampling algorithm the same number of times.

\section{Results}
\label{sec:res}
\subsection{Histograms of Feynman integrals at large loop order}
\label{sec:histos}
We used the methods described in the last section to generate representative samples 
of primitive divergent $\phi^4$ graphs and evaluate their period Feynman integrals at loop orders $8$ to $17$.
The number of graphs we sampled at each loop order is listed in Table~\ref{tab:data}.
We ran the computation in bunches at low priority on the ETH Euler computing cluster. Due to a maintenance event, our computation was interrupted, and some data points were lost.
Hence, the number of samples differs slightly at each loop order.
There is no correlation between the probability of a data point being lost and its value.
At 17 loops, we took fewer samples because we only had limited access to the required large-memory nodes. All our data points of randomly sampled Feynman graphs are available as machine-readable tables in the ancillary material to this article's arXiv version.

Each integral was evaluated using the non-parallelized version of the tropical sampling implementation published with \cite{Borinsky:2020rqs}.  At 17 loops a single integration required about 128 GB of RAM and an average runtime of $1.6 \cdot 10^4$ seconds.  Roughly two-thirds of this time was spent on the pre-processing of the integral, i.e., constructing the tropical measure (see \cite[Sec. 7.2]{Borinsky:2020rqs} for a precise definition), while the remaining third is spent on the Monte Carlo sampling. 
At lower loop order the pre-processing time and memory requirements decrease substantially. 
The ancillary material includes, for each integral, the time spent on both the pre-processing and the sampling steps.

\begin{figure}[h!]
  \centering
\includegraphics[]{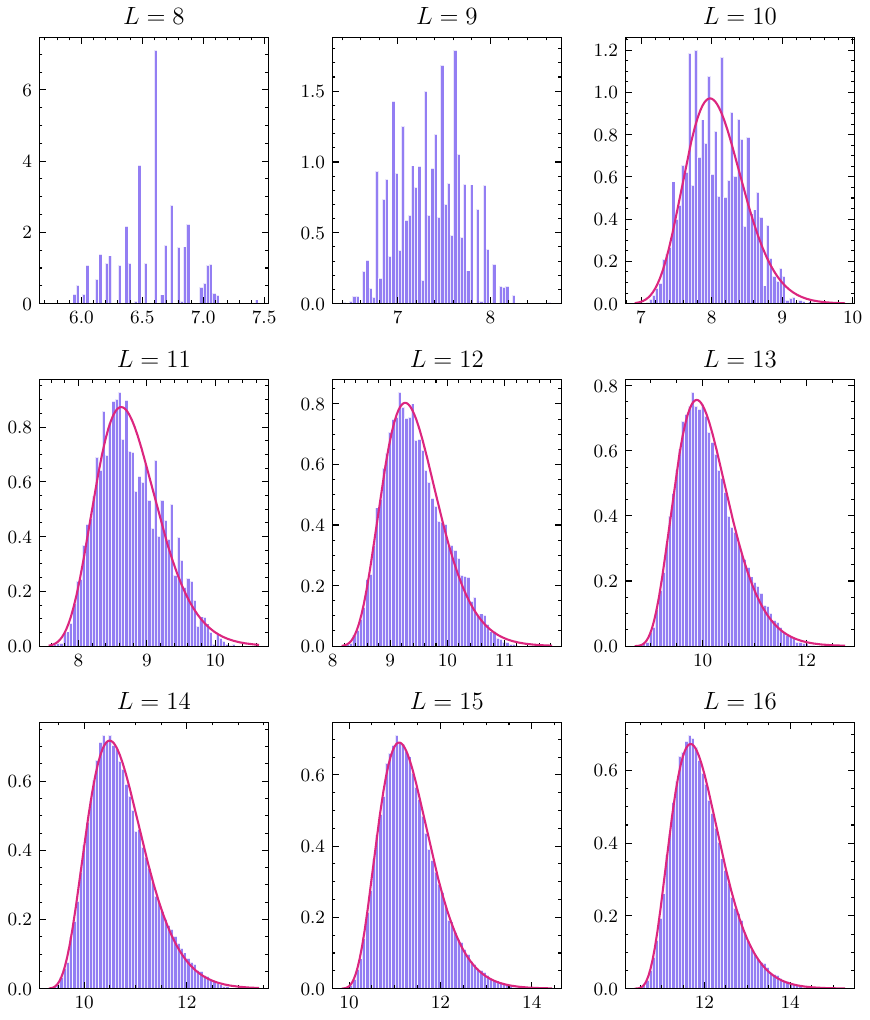}
\caption{Distribution of $\phi^4$ period Feynman integrals up to 16 loops. The $x$-axes show the value of $\log (\mathcal P)$, and the $y$-axes show the density.}
\label{fig:hist2}
\end{figure}

Figure~\ref{fig:hist1} and Figure~\ref{fig:hist2} depict the sampled period Feynman integrals as histograms. 
We evaluated each Feynman graph to $10^{-3}$ relative accuracy using the tropical sampling approach. As this uncertainty is small compared to the statistical uncertainty that stems from the variance of the different Feynman graphs, we can neglect this uncertainty. We confirmed this explicitly by performing our analysis with the uncertainty included and obtaining identical results.

Our data suggests that the distribution of period Feynman integrals is modelled well by the distribution~\eqref{eq:loggamma} for $L\rightarrow \infty$. 
At each loop order, we fitted the parameters $\alpha, \lambda,$ and $\mathcal P_0$ by maximizing the logarithmic \emph{likelihood function}
$$\log \mathcal L = N \left( 
\alpha \log \lambda
- \log \Gamma(\alpha) \right)
+ (\alpha-1) \sum_i \log \log \frac{\mathcal P_i}{\mathcal P_0}
- \lambda \sum_i \log \frac{\mathcal P_i}{\mathcal P_0},
$$
where we sum over all period samples $\mathcal P_1,\ldots, \mathcal P_N$ at a specific loop order.
This likelihood function is readily derived by multiplying 
$N$ copies of the function that multiplies the measure in~\eqref{eq:loggamma} and taking the logarithm.
As the number of samples~$N$ is large, we can estimate the uncertainties of these parameters by approximating the likelihood function near its maximum with a Gaussian.
The resulting parameters with uncertainties are listed in Table~\ref{tab:data}. The uncertainties of the fit parameters were extremely large for $L\in \{8,9\}$. So, we discarded these fits.
The fitted distributions are depicted as red curves in Figure~\ref{fig:hist1} and Figure~\ref{fig:hist2}.

\begin{figure}[h!]
  \centering
  \includegraphics[]{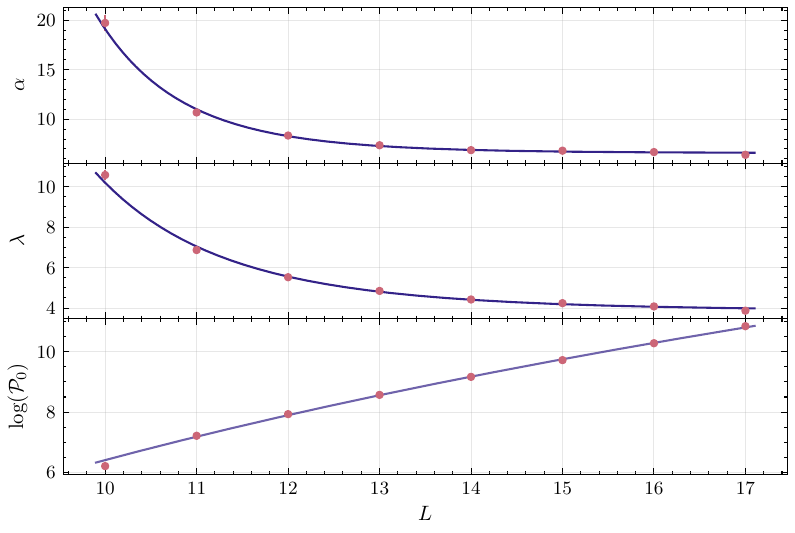}
  \caption{Fitted parameters $\alpha$, $\lambda$, and $\log(\mathcal P_0)$ of the $\log$-$\Gamma$ distribution~\eqref{eq:loggamma} as a function of the loop order $L$. The purple curves are fitted according to the shifted power-law~\eqref{eq:threeparamfit}.}
  \label{fig:params}
\end{figure}

We were not able to give a clear conjecture for the concrete dependence of the parameters $\alpha$, $\lambda$, and $\mathcal{P}_0$ on the loop order $L$. Finding a formula for this explicit dependence, at least in the limit $L\rightarrow \infty$, would be highly beneficial. The dependence is illustrated in Figure~\ref{fig:params}. To give a qualitative assessement, we fitted the parameters $\alpha$, $\lambda$, and $\log \mathcal{P}_0$ using the shifted power-law function:
  \begin{equation}
    \label{eq:threeparamfit}
    f(L; a, b, c) = a L^{b} + c.
  \end{equation}
  The resulting fit parameters $a$, $b$, and $c$ are summarized in Table~\ref{tab:params_fit} 
  , and the corresponding fits are shown as solid curves in Figure~\ref{fig:params}.
Unfortunately, the fits do not provide a clear and sufficiently convincing picture for the dependence in the $L\rightarrow \infty$ limit.

\begin{table}[h]
\centering
{
\begin{tabular}{|r|ccc|}
\hline
 & $a$ & $b$ & $c$ \\
\hline
\rule{0pt}{2.3ex}
$\alpha(L) = f(L; a, b, c)$ & $1.0(1.8) \cdot 10^{12}$ & $-10.92(71)$ & $6.586(77)$ \\
$\lambda(L) = f(L; a, b, c)$ & $1.0(1.2) \cdot 10^{8}$ & $-7.21(51)$ & $3.849(74)$ \\
$\log \mathcal P_0(L) = f(L; a, b, c)$ & $80(67)$ & $0.084(58)$ & $-90(68)$ \\
\hline
\end{tabular}
}
\caption{
  Best-fit parameters $a$, $b$, and $c$ for the $\log$-$\Gamma$ distribution parameters as shifted power-law functions \eqref{eq:threeparamfit} in the loop order $L$.
}
  \label{tab:params_fit}
\end{table}

We checked Conjecture~\ref{conj} quantitatively using Pearson's $\chi^2$ test:
Let $O_i$ be the number of evaluated period Feynman integrals that fall into the $i$-th 
percentile of the distribution~\eqref{eq:loggamma} with the fitted maximum likelihood parameters
at the respective loop order.
The expectation value of the random variable $O_i$ is $N/100$.
So, under the hypothesis that our data follows~\eqref{eq:loggamma}, the quantity $\chi^2 = \frac{100}{N} \sum_i ( O_i - \frac{N}{100} )^2$ is expected to follow a $\chi^2$-distribution with mean $100-3=97$,
as three parameters are fitted.
Table~\ref{tab:data} shows that the ratio $\chi^2/97$ approaches $1$ with increasing loop order, 
consistent with Conjecture~\ref{conj}.
The trend exhibited in Table~\ref{tab:data} is further supported by the computed $p$-value at $L=17$, defined as the probability of obtaining a value of the test statistic $\chi^2$ at least as large as the one observed, assuming the fitted $\log$-$\Gamma$ distribution accurately describes the data. At $L=17$, we find $p=0.087$, indicative of a statistically non-significant deviation from the expected distribution. The observed convergence of the ratio $\chi^2/97$ towards unity further reinforces that deviations diminish progressively with increasing loop order, strongly supporting Conjecture~\ref{conj}.
Figure~\ref{fig:hist2} illustrates how the distribution is approached with increasing loop number,
providing further evidence for Conjecture~\ref{conj}. %

\begin{table}
  \centering
\footnotesize              %
\setlength{\tabcolsep}{2pt}%
\begin{adjustbox}{width=\textwidth}
  \begin{tabular}{|c|rrrrrrrr|}
\hline
$L$ & 
\multicolumn{1}{c}{$\alpha$}&\multicolumn{1}{c}{$\lambda$}&\multicolumn{1}{c}{$\mathcal P_0$}&\multicolumn{1}{c}{$\frac{\chi^2}{97}$}&\multicolumn{1}{c}{$\langle \mathcal{P}(G) \rangle_L^{\rm CLT}$}&\multicolumn{1}{c}{$\langle \mathcal{P}(G) \rangle_L^{\log\text{-}\Gamma}$}&\multicolumn{1}{c}{$\beta_{L+1}^{\rm prim}$}\rule{0pt}{13pt}
& \multicolumn{1}{c|}{$N$}
\\
\hline
$8$\rule{0pt}{2.3ex}
 &
\multicolumn{4}{c}{}&
$7.449(7) \cdot 10^{2}$ &
\multicolumn{1}{c}{}&
$6.064(5) \cdot 10^{6}$ &
$99900$
\\
$9$ &
\multicolumn{4}{c}{}&
$1.644(2) \cdot 10^{3}$ &
\multicolumn{1}{c}{}&
$1.046(1) \cdot 10^{8}$ &
$99800$
\\
$10$ &
$19.72(84)$ &
$10.6(2)$ &
$4.97(19) \cdot 10^{2}$ &
$273.7$ &
$3.504(5) \cdot 10^{3}$ &
$3.522(5) \cdot 10^{3}$ &
$1.890(3) \cdot 10^{9}$ &
$99900$
\\
$11$ &
$10.69(27)$ &
$6.9(1)$ &
$1.36(3) \cdot 10^{3}$ &
$67.7$ &
$7.237(12) \cdot 10^{3}$ &
$7.303(14) \cdot 10^{3}$ &
$3.558(6) \cdot 10^{10}$ &
$99700$
\\
$12$ &
$8.36(17)$ &
$5.53(6)$ &
$2.78(4) \cdot 10^{3}$ &
$19.3$ &
$1.460(3) \cdot 10^{4}$ &
$1.476(3) \cdot 10^{4}$ &
$6.998(14) \cdot 10^{11}$ &
$100000$
\\
$13$ &
$7.38(13)$ &
$4.85(5)$ &
$5.28(6) \cdot 10^{3}$ &
$7.3$ &
$2.869(6) \cdot 10^{4}$ &
$2.900(7) \cdot 10^{4}$ &
$1.429(3) \cdot 10^{13}$ &
$98712$
\\
$14$ &
$6.88(11)$ &
$4.42(4)$ &
$9.58(10) \cdot 10^{3}$ &
$4.6$ &
$5.531(14) \cdot 10^{4}$ &
$5.588(15) \cdot 10^{4}$ &
$3.036(7) \cdot 10^{14}$ &
$100000$
\\
$15$ &
$6.83(10)$ &
$4.24(4)$ &
$1.67(2) \cdot 10^{4}$ &
$2.5$ &
$1.039(3) \cdot 10^{5}$ &
$1.046(3) \cdot 10^{5}$ &
$6.654(18) \cdot 10^{15}$ &
$100000$
\\
$16$ &
$6.688(96)$ &
$4.08(4)$ &
$2.92(3) \cdot 10^{4}$ &
$2.2$ &
$1.910(6) \cdot 10^{5}$ &
$1.916(6) \cdot 10^{5}$ &
$1.504(4) \cdot 10^{17}$ &
$99650$
\\
$17$ &
$6.41(12)$ &
$3.87(5)$ &
$5.14(7) \cdot 10^{4}$ &
$1.2$ &
$3.489(17) \cdot 10^{5}$ &
$3.495(17) \cdot 10^{5}$ &
$3.556(17) \cdot 10^{18}$ &
$44027$
\\
\hline
\end{tabular}
\end{adjustbox}
\caption{Fit parameters $\alpha,\lambda,$ and $\mathcal P_0$ for the period Feynman integral distribution in $\phi^4$ theory, 
the normalized $\chi^2$ value resulting from Pearson's $\chi^2$ test,
the estimated average value of the period Feynman integral
{$\langle {\mathcal P}(G) \rangle_L^{\rm CLT}$ using the central limit theorem,
$\langle {\mathcal P}(G) \rangle_L^{\log \text{-} \Gamma}$ using the $\log$-$\Gamma$ parameters and Eq.~\eqref{eq:moment}}, the 
estimated value of the primitive beta function coefficient,
and the number $N$ of Feynman integrals evaluated at each loop order.
}
\label{tab:data}
\end{table}

\subsection{Estimating moments and the primitive \texorpdfstring{$\phi^4$}{phi4} beta function}

The results of \cite{Balduf:2023ilc} suggest that the moments 
$\langle \mathcal P(G)^k\rangle_L$ diverge for $L \rightarrow \infty$ if $k \geq 2$ (see Table~8 and the discussion before Eq.~(4.8) loc.~cit.). Hence, for sufficiently large loop order, the central limit theorem, which requires a finite second moment, might not be applicable to estimate the value of $\langle {\mathcal P}(G) \rangle_L$. 
So, we can expect a naive sampling approach to fail eventually,
as it will become unfeasible to estimate $\langle {\mathcal P}(G) \rangle_L$
in Eq.~\eqref{eq:betaprim} via averaging over a relatively small amount of samples.
However, under the assumption of the validity of Conjecture~\ref{conj}, we can compute $\langle {\mathcal P}(G) \rangle_L$ for large~$L$ directly using Eq.~\eqref{eq:loggamma}, provided that the parameters $\alpha, \lambda,$ and $\mathcal P_0$ are known:
\begin{align} \label{eq:moment} \langle {\mathcal P}(G)^k \rangle_L = \frac{\lambda^\alpha}{\Gamma(\alpha)} \int_{\mathcal P_0}^\infty \mathcal P^k \left(\log \frac{\mathcal P}{\mathcal P_0}\right)^{\alpha-1} \left(\frac{\mathcal P}{\mathcal P_0}\right)^{-\lambda} \dd( \log \mathcal P)= \mathcal P_0^k \left(\frac{\lambda}{\lambda-k}\right)^\alpha \text{ for large } L. \end{align}
The integral only converges if $\lambda > k$, so outside this range, we find 
$ \langle {\mathcal P}(G)^k \rangle_L=\infty $.

In our data (see Table~\ref{tab:data}),
we observed that $\lambda >3$ for $L\leq 17$.
Hence, up to this loop order, we may assume that 
$\langle \mathcal P(G)^2\rangle_L$ is finite and that we 
can safely estimate $\langle \mathcal P(G)\rangle_L$ using 
the central limit theorem by computing the average of all 
computed values of $\mathcal P(G)$.
The central limit theorem has the advantage over Eq.~\eqref{eq:moment} of providing accurate results 
even when, at a given finite loop order, the empirical distribution deviates slightly from~\eqref{eq:loggamma}.
The uncertainty of this estimate is computed as usual by dividing the sample variance by the number of samples and taking the square root. The results of the estimation are listed in Table~\ref{tab:data} under the column $\langle {\mathcal P}(G) \rangle_L^{\rm CLT}$. 
We also estimated
$ \langle {\mathcal P}(G) \rangle_L $ using Eq.~\eqref{eq:moment}%
, and the results of the estimation are listed in Table~\ref{tab:data} under the column $\langle {\mathcal P}(G) \rangle_L^{\log \text{-} \Gamma}$%
.
The discrepancy between both ways to estimate this expectation value decreases with the loop order $L$. At 17 loops, both methods give confidence intervals that overlap well.

Via \eqref{eq:beta_L}, we can translate our estimates for $\langle \mathcal P(G)\rangle_L$ into the  primitive contribution to the $\phi^4$ beta function.  The necessary values of $Z_L$ were calculated using renormalized 0-dimensional QFT technology \cite[\S 6.3]{Borinsky:2017hkb} or \cite{Cvitanovic:1978wc} (see also \cite{Balduf:2024njk} for a deeper analysis of similar normalization factors from 0-dimensional quantum field theory). Table~\ref{tab:data} also includes the resulting estimates based on the values $\langle {\mathcal P}(G) \rangle_L^{\rm CLT}$ for $\beta_{L+1}^{\rm prim}$. 
These agree with the data for $L\leq 11$ in \cite[Table XIII]{Kompaniets_2017} and confirm estimates in \cite[Table 14]{Balduf:2023ilc} and \cite[Table 4]{Balduf:2024njk} up to 17 of 18 loop orders.

\begin{figure}
  \centering
  \includegraphics[]{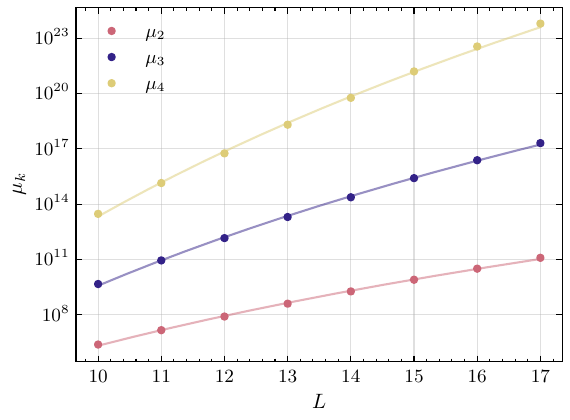}
  \caption{Empirical moments $\mu_k$ of the distribution of $\mathcal P(G)$ as a function of the loop order $L$. The curves are fits to the data points using the function family~\eqref{eq:pL}.}
  \label{fig:moments}
\end{figure}

Extending the results of \cite{Balduf:2023ilc} to the context of Feynman periods sampled proportionally to their symmetry factors, we also empirically investigated the central moments of the distribution of  $\mathcal{P}(G)$ using our dataset. For $k \geq 2$, the central moments $\mu_k$ are defined by 
\begin{align} \label{eq:cMom} \mu_k = \langle \left( \mathcal P(G) - \langle \mathcal P(G) \rangle_L \right)^k \rangle_L. \end{align}
As discussed above, applying the central limit theorem to estimate our quantity of interest, $\langle \mathcal P(G) \rangle_L$, requires that the second moment $\mu_2$ is finite. Further, 
to estimate the accuracy of this estimate itself, the variance of this second moment, $\mathrm{Var}[\mu_2]$, needs to be finite as well. 
Standard arguments in statistics (see, e.g., \cite[Sec.~6.10]{kenney1951mathematics}) 
show 
that finiteness of this variance is implied by the finiteness of the third and fourth moments $\mu_3$ and $\mu_4$.
Hence, we computed the moments $\mu_2,\mu_3,$ and $\mu_4$ from our data.
The results are presented in Table~\ref{tab:moments} and Figure~\ref{fig:moments}.
{Up to a different normalization convention, our results are consistent with those reported in \cite[Table~7]{Balduf:2023ilc}.}

Following the approach of \cite{Balduf:2023ilc}, we fit the computed moments using the function family:
\begin{equation}
\label{eq:pL}
f(L; a, b) = e^a L^b,
\end{equation}
where $a$ and $b$ are fit parameters. The results of these fits are listed in 
Table~\ref{tab:moments_fit} %
and illustrated as colored curves in Figure~\ref{fig:moments}. These fits show that the moments $\mu_k$ rapidly increase with the loop order $L$. Qualitatively, our fit results are also consistent with the behavior reported in \cite{Balduf:2023ilc} for uniformly sampled graphs. 

The rapid growth of these moments suggests that estimating $\langle \mathcal P(G) \rangle_L$ with the central limit theorem will become more and more difficult for larger $L$. Estimating this average using the $\log$-$\Gamma$ hypothesis by fitting the parameters $\alpha,\lambda$ and $\mathcal P_0$ and using formula~\eqref{eq:moment} on the other hand can be expected to stay feasible at arbitrary large loop orders. 
The turning point, where estimating the average using the $\log$-$\Gamma$ hypothesis
gives as accurate results as estimating via the central limit theorem, seems to be at $L\approx 17$ 
as indicated by the data in Table~\ref{tab:data}.

\begin{table}
\centering
\begin{tabular}{|r|ccc|}
\hline
$L$ & $\mu_2$ & $\mu_3$ & $\mu_4$\\
\hline
\rule{0pt}{2.3ex}
$10$ & $2.4121 \cdot 10^{6}$ & $4.6142 \cdot 10^{9}$ & $2.9535 \cdot 10^{13}$ \\
$11$ & $1.4479 \cdot 10^{7}$ & $8.7965 \cdot 10^{10}$ & $1.3856 \cdot 10^{15}$ \\
$12$ & $7.9490 \cdot 10^{7}$ & $1.4266 \cdot 10^{12}$ & $5.6400 \cdot 10^{16}$ \\
$13$ & $3.9723 \cdot 10^{8}$ & $1.9882 \cdot 10^{13}$ & $2.0226 \cdot 10^{18}$ \\
$14$ & $1.8339 \cdot 10^{9}$ & $2.3361 \cdot 10^{14}$ & $5.7832 \cdot 10^{19}$ \\
$15$ & $7.9001 \cdot 10^{9}$ & $2.5503 \cdot 10^{15}$ & $1.5655 \cdot 10^{21}$ \\
$16$ & $3.1370 \cdot 10^{10}$ & $2.3915 \cdot 10^{16}$ & $3.5246 \cdot 10^{22}$ \\
$17$ & $1.2168 \cdot 10^{11}$ & $2.0207 \cdot 10^{17}$ & $6.1323 \cdot 10^{23}$ \\
\hline
\end{tabular}

\caption{Empirically measured moments $\mu_k$ at different loop orders $L$.}
  \label{tab:moments}
\end{table}

\begin{table}[h]
\centering
\begin{tabular}{|r|cc|}
\hline
 & $a$ & $b$ \\
\hline
$\mu_2$ & $-32.47(57)$ & $20.42(22)$ \\
$\mu_3$ & $-54.44(78)$ & $33.22(30)$ \\
$\mu_4$ & $-71.9(1.6)$ & $44.54(63)$ \\
\hline
\end{tabular}
\caption{
  Best-fit parameters $a$ and $b$ for the empirical moments $\mu_k$ using the function family~\eqref{eq:pL}.
  }
\label{tab:moments_fit}  
\end{table}

\subsection{Extrapolating \texorpdfstring{$\beta_{L}^{\rm prim}$}{beta-L-prim} to all loop orders using instanton input}
The coefficients $\beta_{L}^{\rm prim}$ of the primitive contribution to the $\phi^4$ beta function are expected to grow factorially.  In fact, assuming the conjecture that $\beta^{\mathrm{prim}}_{L} \sim \beta^{\mathrm{MS}}_{L} $ for large $L$, instanton computations in scalar quantum field theories~\cite{McKane:1978md,McKane:1984eq,McKane:2018ocs} (see also \S IV.B of \cite{Kompaniets_2017} for details) give a precise prediction for the asymptotic behavior:
\begin{align} \label{eq:beta_asy} \beta^{\mathrm{prim}}_{L} = C \cdot L^{{7}/{2}} \cdot L! \cdot \left( 1 + \bigO\left(\frac{1}{L}\right) \right),  \end{align}
where $C$ equals
$$\frac{144\cdot e^{-\frac{15}{4} - 3 \, \gamma_E}}{\pi^{3/2}  A^6} 
\approx 0.024199,$$
with $\gamma_E$ Euler's constant, and $A$ the Glaisher--Kinkelin constant (Eq.~(21) of \cite{Kompaniets_2017} with $n=1$).

Knowing the precise asymptotic behavior of quantities such as $\beta^{\mathrm{MS}}_{L}$ for $L\rightarrow \infty$ is essential to fruitfully apply resummation techniques while turning  divergent perturbative expansions into trustworthy predictions \cite{LeGuillou:1979ixc,Costin:2020pcj,Kompaniets_2017,Giorgini:2024aer}. Our results give new hard data on this behavior.

Besides the primitive contribution to this asymptotic behavior, \emph{renormalons} are expected to give a factorially growing contribution in the $L\rightarrow \infty$ limit. Computing the renormalon contributions to the $\phi^4$ beta function in the MS scheme and analyzing their asymptotic growth rate seems achievable using, e.g., technology from \cite{Broadhurst:2000dq,Balduf:2021kag,Borinsky:2021hnd,Borinsky:2022knn,Balduf:2025fjp}. Comparing these contributions with the primitive contribution would shed new light on the question of whether instanton or renormalon contributions dominate at large loop order \cite{Beneke:1998ui,DiPietro:2021yxb,Dunne:2021lie,Brezin:2023kyv}. %

Unfortunately, our data for $\beta^{\mathrm{prim}}_{L}$ does not provide compelling evidence for the asymptotic behavior~\eqref{eq:beta_asy}. The main obstacle towards a clear verdict seems not to be the limited accuracy, but the still relatively low loop order of our data points.
We find, via goodness-of-fit estimates, that our data for $\beta^{\mathrm{prim}}_{L}$ is compatible with the following Ansatz, which Eq.~\eqref{eq:beta_asy} directly inspires:
\begin{align} \label{eq:beta_fit} \beta^{\mathrm{prim}}_{L} = L^{{7}/{2}} \cdot L! \cdot \left( c_0 + \frac{c_1}{L} + \frac{c_2}{L^2} + \cdots \right) \text{ as } L\rightarrow \infty\, , \end{align}
where $c_0, c_1, \ldots$ are free parameters. 
Fitting this Ansatz to our data for $\beta^{\mathrm{prim}}_{L}$ results in decent fits.
Fixing the fit parameter $c_0$ to $C$, as defined above, reduces the fit quality dramatically.

By varying the fit range and the cut-off point of the expansion in powers of $1/L$ in~\eqref{eq:beta_fit},
we obtained various estimates for the value of $c_0$. 
With reasonable choices for both the fit range and the cut-off point,
$c_0$ consistently falls into the range of $0.055 \pm 0.015$.
Even though the predicted value of $C\approx 0.024$ lies outside this range,
the large margin of error does not allow for a clear verdict on the validity of~\eqref{eq:beta_asy}.
Moreover, results of Balduf and Th\"urigen~\cite{Balduf:2024njk} show that asymptotic estimates such as ours, which
rely on low-order computations, suffer from various biases. The authors highlight a particular case relevant to $\phi^4$ theory, where the true asymptotic behavior only becomes apparent if the first 25 perturbative coefficients are known. %

\section{Conclusion}
\label{sec:conc}
Empirically, we observed that the value of Feynman integrals converges (weakly) to a specific distribution 
once the loop order gets large. We provided evidence for this by studying the primitive contribution
to the $\phi^4$ beta function in four-dimensional spacetime. The specific limiting distribution 
we find is the $\log$-$\Gamma$ distribution --- a distribution well-known in statistics and probability theory.
We expect limiting distributions of the same or similar shapes to appear in Feynman perturbative expansions
of observables in other quantum field theories.

The limiting distribution has three remaining parameters, $\alpha, \lambda, $ and $\mathcal P_0$, that we fix by fitting at each loop order. It would be highly beneficial to find explicit limiting laws for these three parameters, i.e.~to find the asymptotic behavior of these numbers for $L\rightarrow \infty$. Here, we only studied the behaviour of these parameters qualitatively by making shifted power-law fits.

We gathered large amounts of data on the value of Feynman integrals to come to our conclusions. 
This data is available with the arXiv version of this article. 

We used our data to compute the primitive contribution to the $\phi^4$ beta function up to 17 loops confirming previous results by Balduf \cite{Balduf:2023ilc}.
We discussed the extrapolation of this contribution to all loop orders. It is conjectured that the primitive beta function equals the $\phi^4$ beta function in the minimal subtraction scheme when $L\rightarrow \infty$, so our extrapolation provides a (conjectured) estimate of also this beta function at infinite loop order.
The beta function in the minimal subtraction scheme has numerous phenomenological applications via the Wilson--Fischer approach to critical phenomena.  We postpone the discussions of the phenomenological implications of our findings (e.g., on the critical exponents of the $D=3$ Ising model) to future~work.

Unfortunately, reaching a verdict of full or only partial agreement between our data and predictions from instanton computations is still impossible. 
Our limited data, the resulting poor fit quality, and the large number of sources for numerical perturbations  do not allow a clear conclusion. More data at an even higher loop order seems necessary to complete the picture. The key limiting factor of our computations, which kept us from studying even higher loop orders, was the memory requirements of the tropical sampling implementation~\cite{githubtropsampling}. Harnessing more properties of the Hepp bound from \cite{Panzer:2019yxl} and the (tropical) geometry of Feynman integrals might reduce these requirements and make higher loop orders accessible (see~\cite[\S 8.1 and \S 8.3]{Borinsky:2020rqs}).

\section*{Acknowledgments}
We thank Paul Balduf, Erik Panzer, and Oliver Schnetz for their valuable comments and suggestions for an early version of this paper, Paul Balduf and Erik Panzer for helpful discussions, and Babis Anastasiou for his support.
Research at Perimeter Institute is supported in part by the Government of Canada through the Department of Innovation, Science and Economic Development and by the Province of Ontario through the Ministry of Colleges and Universities. MB was supported by Dr.\ Max Rössler, the Walter Haefner Foundation and the ETH Zürich Foundation. This work was supported by the Swiss National Science Foundation through its project funding scheme (grant number 10001706). We used the ETH Euler cluster for our calculations.

\providecommand{\href}[2]{#2}\begingroup\raggedright\endgroup %

\end{document}